\documentclass[prc,twocolumn,showpacs,amsmath,amssymb]{revtex4}
\usepackage{times}
\usepackage{graphicx}
\usepackage{dcolumn}
\usepackage{bm}
\usepackage{mathrsfs}
\usepackage{amsmath}
\usepackage{textcomp}
\usepackage{amstext}
\begin{document}
\title{The effect of isoscalar-isovector coupling in infinite nuclear matter}
\author{S. K. Singh$^{1}$, M. Bhuyan$^{1,2}$, P. K. Panda$^{3}$, and 
S. K. Patra$^{1}$}
\affiliation{$^1$ Institute of Physics, Sachivalaya Marg, 
Bhubaneswar-751 005, India. \\ 
$^2$ School of Physics, Sambalpur University, Jyotivihar, 
Burla-768 019, India.\\
$^3$ Department of Physics, C. V. Raman College of Engineering, 
Bhubaneswar-752054, India.
}

\date{\today}

\begin{abstract}
Working on the framework of Relativistic Mean Field theory, we exposed 
the effect of nonlinear isoscalar-isovector coupling on G2 parameter 
set on the density dependence of nuclear symmetry energy in infinite 
nuclear matter. The observables like symmetric energy and few related 
coefficients are studied systematically. We presented the results of 
stiff symmetry energy at sub-saturation densities and a soft variation 
at normal densities. Correlation between the symmetric energy and the 
isoscalar-isovector coupling parameter fully demonstrated for wide 
range of density. The work further extended to the octet system and 
showed the effect of coupling over the equation of state. 
\end{abstract}


\maketitle

\section{Introduction}

The best possible and well defined theoretical laboratory to studied 
for many body system is infinite nuclear matter at certain conditions. 
To review the status of microscopic studies of nuclear matter and neutron 
rich matter, that already reaches to the destination by mean-field models 
(relativistic and non-relativistic) \cite{lala97,vretenar00,vretenar03,
colo92,blaizot95,hama97} and although by some other-microscopic, methods 
such as Brueckner-Hartree-Fock (BHF) \cite{bruk54,bethe56,gold57,bethe57} 
and Dirac-Brueckner-Hartree-Fock (DBHF) \cite{ana83,horo84,horo87,brock65,
brock84}. The isospin and density dependence of the nuclear 
symmetry energy, $E_{sym}$, is the current interest because of its 
implications not only in nuclear physics but also in astrophysics. The 
softening in equation of state (EOS) of nuclear matter likely lead 
to an exciting problem in astro-nuclear physics from few decades to 
till date. Mean while the novel phenomena like formation of superheavy 
nuclei in astrophysical system is the front learning about the island 
of stability for that regions. 

The prediction of compressional modulus ($K_0$) in symmetric nuclear 
matter for nonrelativistic and relativistic models differ by about 
25 in magnitude. In other word, the nonrelativistic models predict 
$K_0\sim$ 220-235 MeV \cite{colo92,blaizot95,hama97}, relativistic 
models argue for $K_0\sim$ 250-280 MeV \cite{lala97,vretenar00,vretenar03}. 
The density dependence of the symmetry energy $E_{sym}$ is not exposed 
fully at present, which may be responsible for the above discrepancy 
\cite{pikar02}. Note that the $E_{sym}$ is the difference between the 
energy of pure neutron matter and normal nuclear matter, which is a 
tremendous approximation \cite{vretenar03,pikar04,agarwal03}. The large 
value of $K_0$ shows appropriately stiff symmetric energy i.e. rises 
rapidly with baryon density \cite{pikar02,pikar04}. This make 
a passing reference between $K_0$ and $E_{sym}$, which is a function of 
density and isospin component of scalar and vector mesons.
Thus, it is enlightening to explore the effects of $E_{sym}$ 
in the RMF model, which was limited to a narrow range from the 
analysis of skin data, on the composition and structure of hot 
proton neutron and cold neutron star matter that hold a large 
density range \cite{glender92,heisen99,aru04,panda97,casare98}.

In this work, we introduce an extra term on the top of the $G1$ 
or $G2$ forces to the Lagrangian, which is  
the combined effect of scalar and vector field with isoscalar-
isovector coupling constant ($\Lambda_v$). The newly added term 
affect to the symmetric energy and the slope parameter $L_{sym}$. 
The problem is the stiffness of the $E_{sym}$ with respect to 
baryonic density followed a softer path with respect to density along 
with this extra added coupling. Without this additional coupling 
constants one may not be overcome the hindrance in the nuclear matter. 
The most important prospectus of the coupling constant is that the 
various other properties and nuclear matter observables are not 
seriously compromising in their magnitude.  

\section{The relativistic mean field (RMF) theory}

The complete description of EOS can be implemented by using quantum 
chromodyanmics (QCD), which is the fundamental theory of the strong 
interaction. The concept of effective field theory (EFT) is applicable 
to low energy QCD \cite{wein79} and is called as quantum hadrodynamics 
(QHD) \cite{waleka74,waleka86,rein89,serot92,serot97}. The description 
of normal nuclear matter properties \cite{shapi83,waleka74} and finite 
nuclei \cite{horo81,boguta77,gambhir90,ring96} are successfully explained 
by QHD. The exchange of mesons like iso-scalar scalar $\sigma$, 
iso-scalar vector $\omega$, iso-vector vector $\rho$ and the pseudoscalar 
$\pi$ are responsible for nucleon interaction and increases the degree 
of freedom. The nucleons are considered as Dirac spinors moving in 
classical meson fields. In mean-field approximation, the contribution 
of the $\pi$ meson is zero, because of its pseudospin nature. The 
extension of the standard relativistic mean-field (RMF) theory with 
addition of non-linear scalar-vector and vector-vector self interaction 
is called as E-RMF proposed by {\it Furnstahl}, {\it Serot}, and 
{\it Tang} \cite{furn96,furn97} (see also {\it Miller} and {\it Serot} 
\cite{muller96}. The coupling constants are self-consistent with the 
underlying symmetries of QCD by non-renormalization. The concept of 
naturalness can focus by expanding the non-linear Lagrangian and organized 
it in increasing power of field and their derivative according to their  
dimensional analysis \cite{geo84,geo93} and truncated at a given level 
of accuracy \cite{rusnak97,furn00,serot04}. Here we have taken the fourth 
order of expanded field and in addition to this we have taken the special 
interaction between the $\omega$ and $\rho$ with a new coupling constant 
$\Lambda_v$ to get good result in the study of $E_{sym}$ and other 
co-efficients. The modified Lagrangian is given as \cite{bedn01,bedn06}: 

\begin{eqnarray}
{\cal L}&=&\sum_B\overline{\psi}\left(
\{i\gamma^{\mu}D_{\mu}-m_B+g_{\sigma}\sigma\right)\psi \nonumber \\
&& + \frac{1}{2}\partial_{\mu}\sigma\partial_{\mu}\sigma-m_{\sigma}^2\sigma^2
\left(\frac{1}{2}+\frac{\kappa_3}{3!}\frac{g_{\sigma}\sigma}{m_B}
+\frac{\kappa_4}{4!}\frac{g_{\sigma}^2\sigma^2}{m_B^2}\right) \nonumber \\
&& - \frac{1}{4}\Omega_{\mu\nu}\Omega^{\mu\nu}+\frac{1}{2}m_{\omega}^2
\omega_{\mu}\omega^{\mu}\left(1+\eta_1\frac{g_{\sigma}\sigma}{m_B}
+\frac{\eta_2}{2}\frac{g_{\sigma}^2\sigma^2}{m_B^2}\right) \nonumber \\
&& -\frac{1}{4}R_{\mu\nu}^aR^{\mu\nu a}+\frac{1}{2}m_{\rho}^2
\rho_{\mu}^a\rho^{a\mu}\left(1+\eta_{\rho}
\frac{g_{\sigma}\sigma}{m_B} \right) \nonumber \\
&& +\frac{1}{4!}\zeta_0 g_{\omega}^2\left(\omega_{\mu}\omega^{\mu}\right)^2
+\Lambda_v(g_{\rho}^2\rho_{\mu}^a\rho^{\mu a})
(g_{\omega}^2\omega_{\mu}\omega^{\mu})
\end{eqnarray}

The co-variant derivative $D_{\mu}$ is defined as
\begin{eqnarray}
D_{\mu}=\partial_{\mu}+ig_{\omega}\omega_{\mu}
+ig_{\rho}I_3\tau^a\rho_{\mu}^a,
\end{eqnarray}
where $R_{\mu\nu}^a$ and $\Omega_{\mu\nu}$ are field tensors and 
defined as follow 
\begin{eqnarray}
R_{\mu\nu}^a=\partial_{\mu}\rho_{\nu}^a-\partial_{\nu}\rho_{\mu}^a
+g_{\rho}\epsilon_{abc}\rho_{\mu}^b\rho_{\nu}^c,
\end{eqnarray}

\begin{eqnarray}
\Omega_{\mu\nu}=\partial_{\mu}\omega_{\nu}-\partial_{\nu}\omega_{\mu},
\end{eqnarray}

Here $m_B$ is the mass of nucleon, and $m_{\sigma}$, $m_{\omega}$ and 
$m_{\rho}$ are the masses of corresponding mesons. The equation of 
motion derived from the Lagrangian using mean-field approximation 
self-consistently. The field equation for different meson fields 
are given by
\begin{eqnarray}
m_{\sigma}^2\left(\sigma_0+\frac{g_{\sigma}\kappa_3\sigma_0^2}{2m_B^2}
+\frac{g_{\sigma}^2\kappa_4\sigma_0^3}{6m_B^2} \right) \nonumber \\
-\frac{1}{2}m_{\omega}^2\left(\eta_1\frac{g_{\sigma}}{m_B}
+\eta_2\frac{g_{\sigma}^2\sigma_0}{m_B^2}\right)\omega_0^2 \nonumber \\
-\frac{1}{2}m_{\rho}^2\eta_{\rho}\frac{g_{\sigma}\rho_0^2}{m_B}
=\sum g_{\sigma}\rho_s,
\end{eqnarray} 
\begin{eqnarray}
m_{\omega}^2\left(1+\eta_1\frac{g_{\sigma} \sigma_0}{m_B}
+\eta_2\frac{g_{\sigma}^2\sigma_0^2}{m_B^2}\right)\omega_0 \nonumber \\
-\frac{1}{6}\zeta_0g_{\omega}^2\omega_0^3 
+ 2\Lambda_v g_{rho}^2\rho_0^2g_{\omega}^2\omega_0
=\sum g_{\omega}\rho_B,
\end{eqnarray}
\begin{eqnarray}
\left[m_{\rho}^2\left(1+\eta_{\rho}\frac{g_{\sigma}\sigma_0}{m_B}\right)
+2\Lambda_v g_{rho}^2g_{\omega}^2\omega_0^2\right]\rho_0
=\frac{1}{2}\sum g_{\rho}I_3\rho
\end{eqnarray}
The scalar ($\rho_s$) and vector ($\rho$) densities are given as 
follow:
\begin{eqnarray}
\rho_s=\frac{1}{\pi^2}\int_0^{k_B}\frac{k^2dk}{E_B^*}
\end{eqnarray}
and 
\begin{eqnarray}
\rho_B=\frac{1}{\pi^2}\int_0^{k_B}k^2dk
\end{eqnarray}
where, $E_B^*=\sqrt(k^2+m_{B^{*2}})$ is the effective energy, $k_B$ is 
the Fermi momentum, $m_B^*=m_B-g_{\sigma}\sigma$ stands for effective 
mass (which solved self-consistently), $I_3$ is the third component of 
isospin projection and $B$ stands for baryon. The energy and pressure 
density depends on the effective mass $m_B^*$ of the system, so first 
need to solve this self-consistent Eqn. (4). As the self-consistent 
field obtained, the expression for energy and pressure density are 
given as;
\begin{eqnarray}
\mathscr{E}&=&\sum_B\frac{2}{(2\pi)^3}\int_0^{k_B}d^3kE_B^*(k)
+\frac{1}{8}\zeta_0g_{\omega}^2\omega_0^4 \nonumber \\ 
&& + m_{\sigma}^2\sigma_0^2\left(\frac{1}{2}+\frac{\kappa_3}{3!}
\frac{g_{\sigma}\sigma}{m_B}+\frac{\kappa_4}{4!}
\frac{g_{\sigma}^2\sigma^2}{m_B^2}\right) \nonumber \\
&& + \frac{1}{2}m_{\omega}^2 \omega_0^2\left(1+\eta_1
\frac{g_{\sigma}\sigma}{m_B}+\frac{\eta_2}{2}
\frac{g_{\sigma}^2\sigma^2}{m_B^2}\right) \nonumber \\
&& + \frac{1}{2}m_{\rho}^2 \rho_0^2\left(1+\eta_{\rho}
\frac{g_{\sigma}\sigma}{m_B} \right) \nonumber \\ 
&& + 3\Lambda_v\Omega_0^2 R_{0}^2+\sum_l\varepsilon_l,
\end{eqnarray}   

and 

\begin{eqnarray}
\mathscr{P}&=&\sum_B\frac{2}{3(2\pi)^3}\int_0^{k_B}d^3kE_B^*(k)
+\frac{1}{8}\zeta_0g_{\omega}^2\omega_0^4 \nonumber \\ 
&& + m_{\sigma}^2\sigma_0^2\left(\frac{1}{2}+\frac{\kappa_3}{3!}
\frac{g_{\sigma}\sigma}{m_B}+\frac{\kappa_4}{4!}
\frac{g_{\sigma}^2\sigma^2}{m_B^2}\right) \nonumber \\
&& + \frac{1}{2}m_{\omega}^2 \omega_0^2\left(1+\eta_1
\frac{g_{\sigma}\sigma}{m_B}+\frac{\eta_2}{2}
\frac{g_{\sigma}^2\sigma^2}{m_B^2}\right) \nonumber \\
&& + \frac{1}{2}m_{\rho}^2 \rho_0^2\left(1+\eta_{\rho}
\frac{g_{\sigma}\sigma}{m_B} \right) \nonumber \\
&& + \Lambda_v\Omega_0^2 R_{0}^2+\sum_l P_l,
\end{eqnarray}
where $P_l$ and $\varepsilon_l$ are lepton's pressure and energy density,
respectively (and here we have taken electron only). Note that the 
following quantities are redefined as, $\Omega_0 \equiv g_{\omega}\omega_0$ 
and $R_0 \equiv g_{\rho} \rho_{03}$. At zero temperature, the pressure 
of the system obtained from energy density and its first derivative, 
which is thermodynamically consistent in mean-field (MF) approximation. 


\section{Calculations and Discussion}
We have solved the mean field equations self-consistently and 
parametrized the coupling constant by renormalization. We have 
made attention to G2 \cite{furn96,furn97} parametrization sets 
of the E-RMF model to reproduce the fundamental properties of 
nuclear matter. In this parameter set $\zeta$ coupling constant 
is determined by self-interactions of $\omega$-meson, which is 
responsible for the softening the EOS at high density and reproduced 
the maximum mass of neutron star. For softening the symmetric energy 
of the system at nuclear matter density, we have added an extra coupling 
between iso-scalar and iso-vector field with coupling constant 
$\Lambda_v$. This made make possible to soften the $E_{sym}$ and 
reproduced saturation properties for the force parameter at $\Lambda_v$ 
$\sim$ 0.00. In other word, $\Lambda_v$ is simply a self interaction 
constant between isoscalar and isovector field, which does not change 
any other nuclear matter properties like energy density, pressure density 
and so many considerably.  


\subsection{Energy and Pressure density}
Infinite nuclear matter is important for the investigation of physical 
quantities relevant to heavy nuclei and compact objects like neutron 
star. At saturation density $\rho_0$, the binding energy per particle 
$\epsilon$ as a function of density is well established quantities 
consistent to thermodynamic relation at certain temperature. These 
translational invariance of the system facilitates theoretical 
calculations of these observables. The prediction of E-RMF 
(G2+$\Lambda_v$) for EOS of nuclear matter (NM) (solid black line) 
compared with DBHF \cite{dbhf90} and other theoretical predictions 
\cite{shf12} are in Fig. 1 (a). From the figure, it is clear that 
all the prediction well matches with respect to density. Fig. 1(b), 
shows the comparison of pressure density with experimental data 
\cite {dan02} and theoretical predictions \cite{fran10,dbhf90,typel05}. 
The obtained results for energy density (see Fig. 1) from Ref. 
\cite{aru04} is exactly same to the present calculation. This implies 
that the EOS of SNM is not affected by the new coupling with $\Lambda_v$. 
Being insensitive to this observable, the aim of this present work to 
pursue the systematic variation of $E_{sym}$ by employing the 
isoscalar-isovector coupling parameter $\Lambda_v$, which discussed 
below. 
    
\begin{figure}
\vspace{0.6cm}
\begin{center}
\includegraphics[width=1.0\columnwidth]{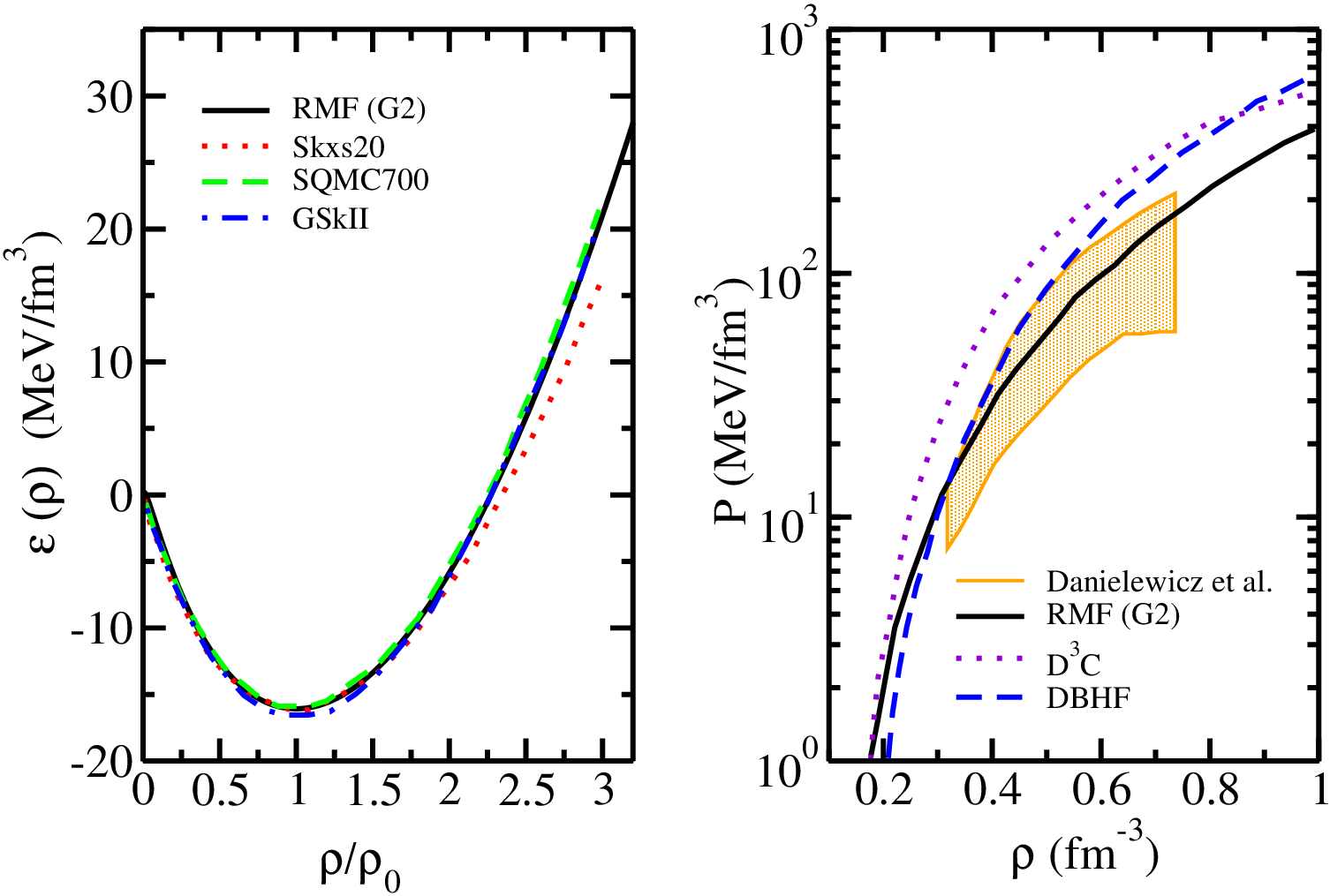}
\caption{\label{epn}The EOS or energy per nucleon and pressure density 
as a function of density from RMF compare with the other predictions. 
}
\end{center}
\end{figure}

\begin{figure}
\vspace{0.6cm}
\begin{center}
\includegraphics[width=0.9\columnwidth]{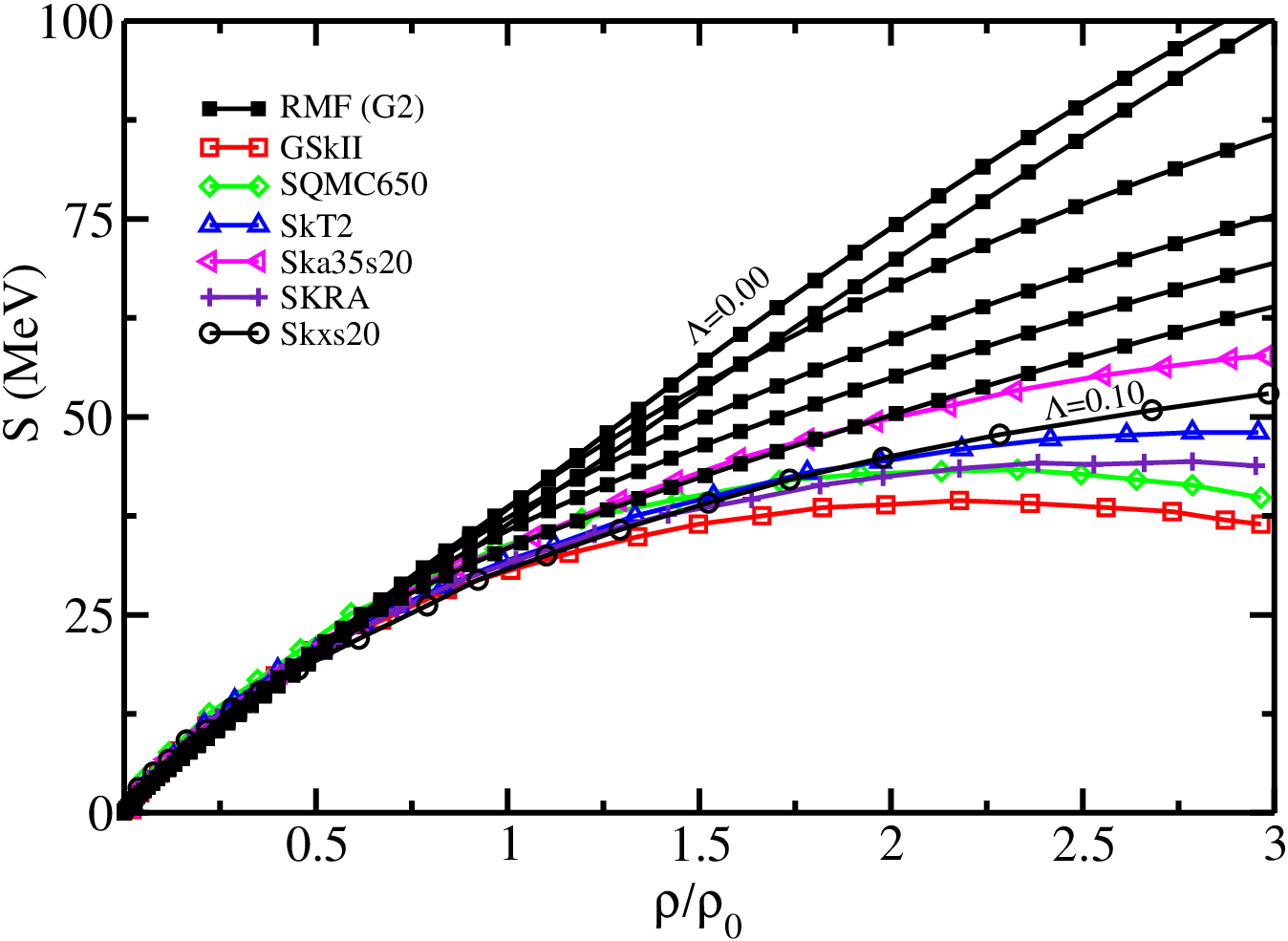}
\caption{\label{sym}The results of symmetric energy $E_{sym}$ 
is a function of density $\rho$ from RMF for different values 
of $\Lambda_v$ compare with others predictions.
}
\end{center}
\end{figure}

\begin{figure*}
\vspace{0.6cm}
\begin{center}
\includegraphics[width=1.8\columnwidth,height=1.0\columnwidth]{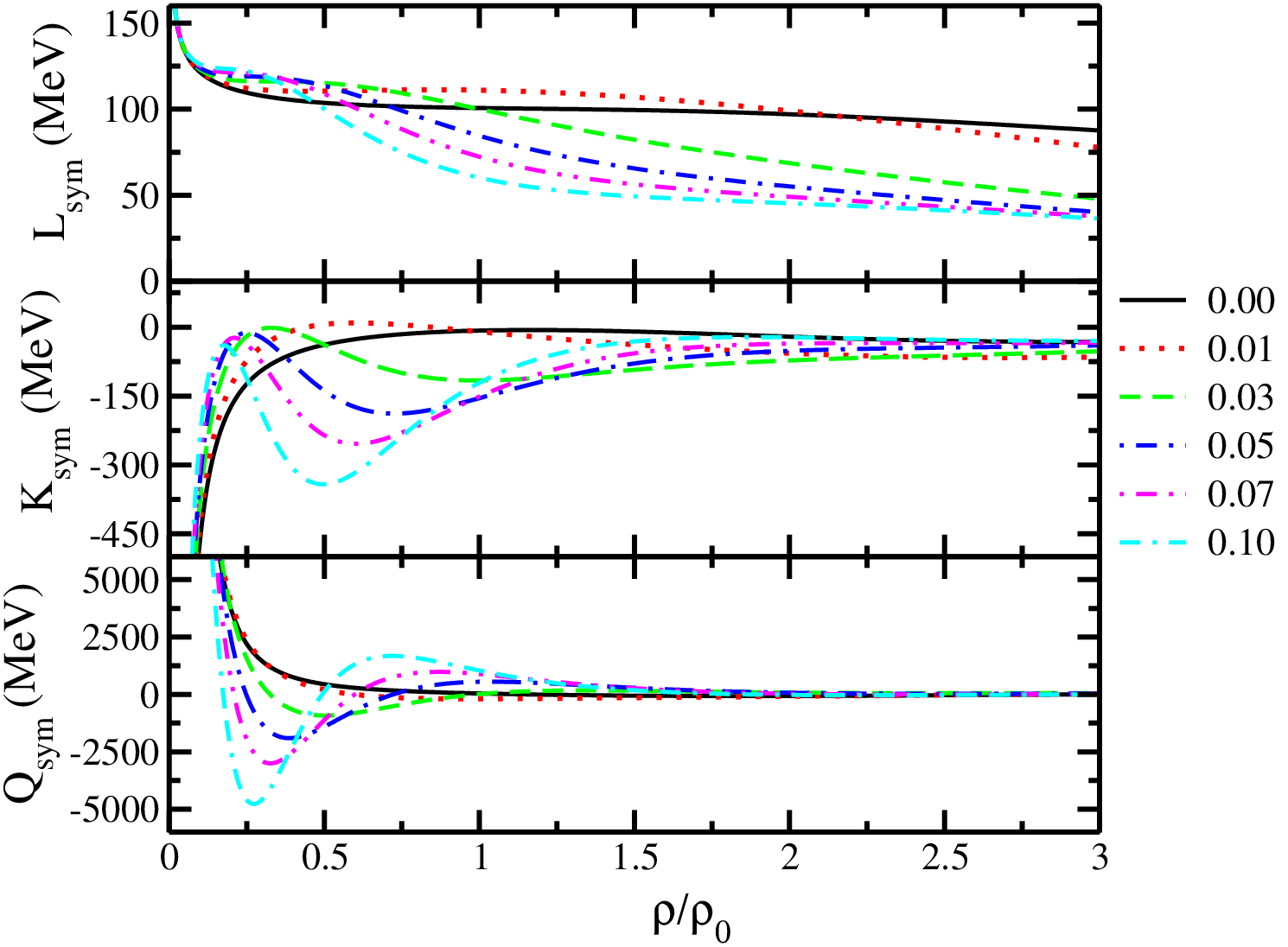}
\caption{\label{kl}The slope parameter $L_{sym}$ along with $K_{sym}$ 
and $Q_{sym}$ for different value of $\Lambda_v$, with $G2$ parameter set.}
\end{center}
\end{figure*}


\subsection{Symmetry Energy }
The symmetric energy $E_{sym}$, which is important in infinite 
nuclear matter and finite nuclei because of isospin dependent in 
the interaction. The expression of this key observable is defined 
as

\begin{eqnarray}
E_{sym}(\rho)=\frac{k_F^2}{6E_F^*}+\frac{g_{\rho}^2\rho_B}{2}
\left(\frac{1}{4{m_{\rho}^{*2}}}+\frac{\Lambda_vg_{\rho}^2
g_{\omega}^2\omega^2}{m_{\rho}^{*2}}\right),
\end{eqnarray}
where the effective energy $E_F^*=\sqrt(k_F^2+m^{*2})$, $k_F$ is the 
Fermi momentum and the effective mass $m^*=m-g_S\phi_0$. The effective 
mass of the $\rho$-meson modified because of nonlinear coupling 
($\rho-\omega$) interaction and is given by
\begin{eqnarray}
m_{\rho}^{*2}=\left(1+\eta_{rho}\frac{g_{\sigma}\sigma}{m_B}\right)m_{\rho}^2
+2g_{\rho}^2(\Lambda_vg_v^2\omega_0^2),
\end{eqnarray}
where $m_{\rho}$ is the mass of the $\rho$-meson. It is noted that 
the non-linear isoscalar-isovector coupling ($\Lambda_v$) modified 
the density dependent of $E_{sym}$ without affecting the saturation 
properties of the SNM (as shown in Fig. 1 (a)). In this model the 
symmetric nuclear matter saturates at a Fermi momentum of $k_F$ 
= 1.00 $fm^{−1}$ with $E\sim$ 16.0 MeV and an compressibility 
of $K_0$ = 215 MeV \cite{aru04}. We can expand the density dependence 
of the $E_{sym}$ with the help of these $3$-co-efficients as a function of
${\cal Y}=\left(\rho-\rho_0\right)/3\rho_0$ as follow:
\begin{eqnarray}
{\cal S}(\rho)&=&E_{sym}(\rho_0)+L_{sym}{\cal Y}
+\frac{1}{2}K_{sym}{\cal Y}^2 \nonumber \\
&& +\frac{1}{6}Q_{sym}{\cal Y}^3 +O({\cal Y}^4),
\end{eqnarray}
where $L_{sym}$ and $K_{sym}$ are defined as  
\begin{eqnarray}
L_{sym}=3\rho_0\left(\frac{\partial E_{sym}}{\partial \rho}
\right)_{\rho=\rho_0},
\end{eqnarray}
 
\begin{eqnarray}
K_{sym}=3\left(\rho_0\frac{\partial L_{sym}}{\partial\rho}
- L_{sym}\right)_{\rho=\rho_0}.
\end{eqnarray}
and 
\begin{eqnarray}
Q_{sym}=27\rho_0^3\left(\frac{\partial^3 E_{sym}}{\partial \rho^3}
\right)_{\rho=\rho_0}.
\end{eqnarray}
Here $L_{sym}$ is called the slope parameter, which is nothing but 
the slope of $E_{sym}$. The quantity $K_{sym}$ represents as the 
curvature of $E_{sym}$ with respect to density. In SNM, the value of 
$K_{sym}$ at saturation density ($\rho_0$) entitled by $K_0$ and called 
as symmetry incompressibility. Finally, $Q_{sym}$ is the third derivative 
of the symmetric energy, which called as skewness coefficient \cite{shf12}.
As shown in the Fig. 2, we plotted the $E_{sym}$ for RMF (G2+$\Lambda_v$) 
compare with other theoretical predictions \cite{shf12}. The softness of        
the $E_{sym}$ with the coupling $\Lambda_v$ clearly observed from the 
figure. Further, the rate of change of $E_{sym}$ i.e. the slope parameter 
$L_{sym}$ for different values of $\Lambda_v$ is displayed in Fig. 3. 
Both the figure shows similar behaviour with the coupling constant and 
with baryonic density.  

There are many references which trying to fix the constraint on 
symmetric energy and $L_{sym}$ \cite{xu10,new11,ste12,latt12,
shf12,tsang12,fatt12}. Finally we comes to the constraint on 
$E_{sym}$ and $L_{sym}$ at saturation density ($\rho_0$) and the 
obtained results are compared with with experimental data at HIC 
\cite{hic09}, PDR \cite{pdr07,pdr10}, IAS \cite{ias} and FRDM 
\cite{frdm12} in Fig. 4 and Table 2. From the figure, it is clear that 
for $\Lambda_v$=0.00, our observables are not within the experimental 
constraint region. For the values of $\Lambda_v$=0.01-0.04, these comes 
within limit and again go out with further increasing the value of 
$\Lambda_v$ =0.05. Hence, from this behavior, we may suggest the 
limiting value of $\Lambda_v$ is 0.01-0.04, which can improved the 
observables with G2 force parameter. 
 
\begin{table}
\renewcommand{\tabcolsep}{0.8cm}
\renewcommand{\arraystretch}{1.4}
\caption{\label{tab:tb1}The results for slope co-efficient $S_{sym}$, $K_{sym}$
for different value of $\Lambda_v$ are in list.}
\begin{tabular}{ccccccccccccccccc}
\hline
$\Lambda_v$ &  $E_{sym}$ & $L_{sym}$ \\
\hline
0.00 & 36.40 & 105.67 \\
0.01 & 38.41 & 111.69 \\
0.02 & 38.87 & 113.09 \\
0.03 & 38.62 & 112.32 \\   
0.04 & 38.02 & 110.52 \\
0.05 & 37.27 & 108.27 \\
0.06 & 36.46 & 105.85 \\   
0.07 & 35.64 & 103.42 \\
0.08 & 34.86 & 101.05 \\
0.09 & 34.11 & 98.79 \\
0.10 & 33.39 & 96.65 \\
\hline
\end{tabular}
\end{table}

\begin{figure}
\vspace{0.6cm}
\begin{center}
\includegraphics[width=1.0\columnwidth]{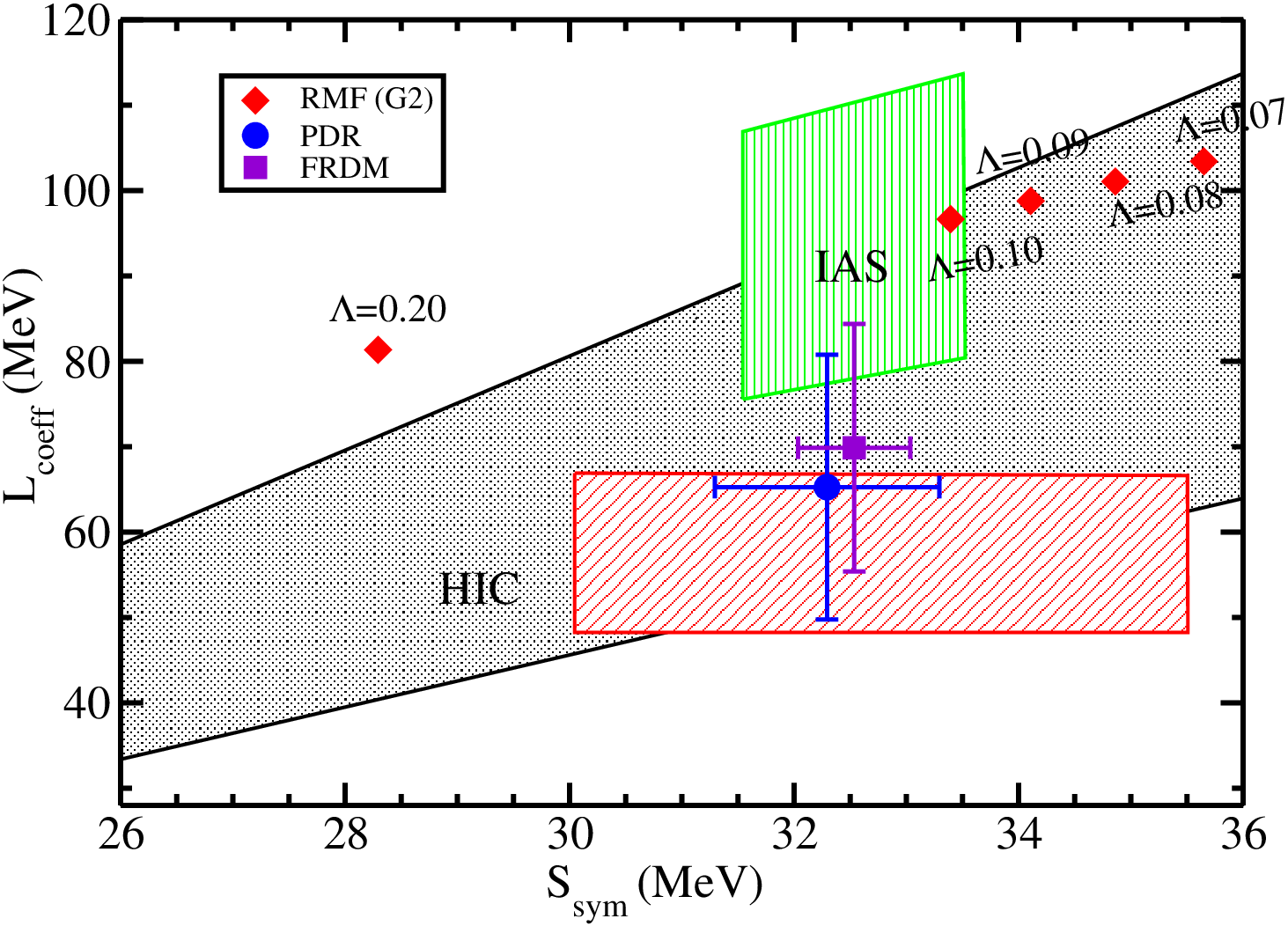}
\caption{\label{flow}The Symmetric energy $S_0$ with respect to 
$L_{sym}$ at saturation density compare with flow data. 
}
\end{center}
\end{figure}

\section{Hadron Matter}
The density at the interior core of the neutron star is very high 
$\sim$ 7-8$\rho_0$ so other hadronic part \cite{bedn01,bedn06,gle97,
gle85,sch99} also come to play role on the equation of state. In this 
section, we extended our calculations to the system with octet 
($n$, $p$, $\Lambda^0$,$\Sigma^{0,\pm1}$, $\Xi^{-,0}$). The purpose 
of this work to show the effect of isoscalar-isovector interaction 
with coupling constant $\Lambda_v$ on the equation of state of the 
nuclear system with full octet. It is worthy to mention that, many authors 
\cite{furn00, aru04, sharma07} have been used $G1$ and $G2$ 
forces for the description of finite and infinite nuclear matter 
properties. Moreover, the non-linear isoscalar-isovector interaction 
in mean field was discussed in Refs. \cite{fsu, fatto10}, here we 
suppose to check this interaction with G2 along with $\Lambda_v$, 
keeping all saturation constraints. We assume that all the hyperon 
in the octet system have the same coupling ratio with mesons. This 
is given by a constant factor which is equal to $\sqrt{2/3}$ 
\cite{sharma07}, the density of the system $\sim$ 7-8 $\rho_0$ and 
assuming that the stars are the composite system of baryons. The 
electron $e^{-1}$ and muon $\mu$ are for maintaining the charge 
neutrality and $\beta-$ equilibrium condition \cite{sharma07,ste02} 
under  the week interactions: 
\begin{equation}
B_1\rightarrow B_2+l+\overline{\nu_l}; \nonumber \\
\quad\quad B_2+l\rightarrow B_1+\nu_l,
\end{equation}
where $B_1$, $B_2$, $l$, $\nu$, $\overline{\nu}$ are baryons, leptons,
neutrino and antinuetrino respectively. In case of octet system, we are 
dealing with neutron star (real system), which became unstable with 
small change in the system. The equation of state obtained for nuclear 
matter with nucleon and hadrons for $G2+\Lambda_v$ is shown in Fig 5. 

We plotted pressure($fm^{-4}$) density with respect to energy density
($fm^{-4}$) in Fig. 5. Here we can see that at high density octet EOS 
becomes stiffer by increasing the $\Lambda_v$ value but it behave different 
way at low density region. Equation of state for necleon case (EOS-NP) 
becomes softer with increasing $\Lambda_v$ value over whole the density 
region. We compare our results with the empirical EOS obtained by 
Steiner {\it et al.} \cite{stei10}. Along with the octet we also plotted 
nucleon EOS at different $\Lambda_v$ values.Nucleon EOS well matches with 
the data of Steiner {\it et al.} \cite{stei10} at low and high density 
but the octet EOS is deviates at high density. The octet EOS compares well 
with the experimental data only at intermediate density. An impressive 
observation is noted from this plot, the influence of $\Lambda_v$ with 
G2 parameter set is not much because in G2 already taken the four power 
of iso-scalar meson so it is more effective then iso-scalar and iso-vector 
interaction.  
\begin{figure}
\vspace{1.5cm}
\begin{center}
\includegraphics[width=1.0\columnwidth]{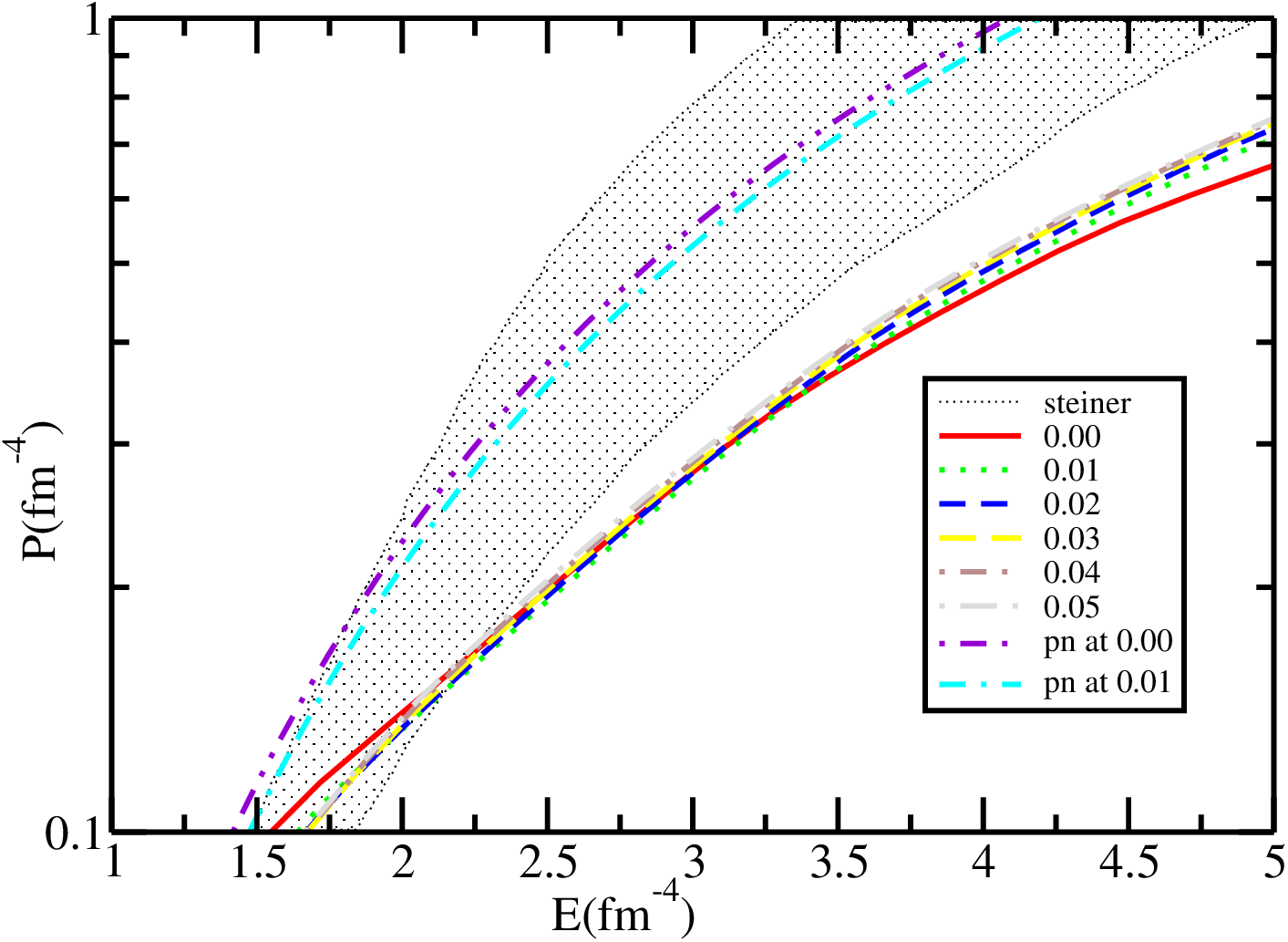}
\caption{\label{oc}Equation of states(EOS) with the E-RMF model for octet and NP 
at different $\Lambda_v$ value.}
\end{center}
\end{figure}

\section{Summary and Conclusions}
After analyzing the result whatever got from this calculation, 
we forcefully say that $\Lambda_v$ i.e. the coupling  
of isoscalar and isovector interaction is very crucial to softening 
the equation of state of nucleonics and octet system. One can easily 
observed that the $E_{sym}$ curve with respect to $\rho$/$\rho_0$ 
became softer with varying $\Lambda_v$ but after $\Lambda_v$ $\sim$ 
$0.05$ it became constant. Same result we got in $L_{sym}$ which 
is the slope of symmetric energy. For symmetric nuclear matter isoscalar 
and isovector interaction do not effect the $E/A$ and pressure density 
curve with respect to $\rho$/$\rho_0$. We got identical curve as in 
Ref. \cite{aru04} for $G1$ and $G2$ sets with experimental data 
\cite{dan02}. In case of hadron, the extra term containing $\Lambda_v$ 
plays an important role in softening the equation of state. 
Hence, we may say that this interaction term is crucial for both cases 
(nucleon and hadron). Further, we extend our calculation to the neutron 
star properties in near future. 

\section{Acknowledgments}
One of the author MB thanks to Institute of Physics for local hospitality 
at the time of work. We thanks to B. K. Sharma for important discussions 
related to this work. 


\end{document}